\documentclass[12pt]{article}
\usepackage{geometry}                
\usepackage{graphicx}
\usepackage{subfig}
\usepackage{floatrow}

\def\ten{\mathcal T}
\def\mass{\mathcal M}
\def\tension{\mathcal T}

\def\s{\mathcal S}

\def\gzero{\hat g_{ab} }

\def\gtt{\gamma_t{}^t}

\def\gii{\gamma_i{} ^i}
\def\grr{\gamma_r {} ^r }

\begin{document}

\begin{titlepage}
\vfill
\begin{flushright}
NSF-KITP-13-081
\end{flushright}

\vskip 3.0cm
\begin{center}
\baselineskip=16pt
{\Large\bf Sum Rule for the ADM Mass and Tensions}
\vskip 0.3cm
{\Large\bf  in Planar AdS Spacetimes}
\vskip 10.mm
{\bf Basem Mahmoud El-Menoufi\footnote{bmahmoud@physics.umass.edu}, Benjamin Ett\footnote{bett@physics.umass.edu}, David Kastor\footnote{kastor@physics.umass.edu},
 Jennie Traschen\footnote{traschen@physics.umass.edu}} 

\vskip 0.5cm
Department of Physics, University of Massachusetts, Amherst, MA 01003
\vspace{6pt}
\end{center}
\vskip 0.2in
\par
\begin{center}
{\bf Abstract}
 \end{center}
\begin{quote}

An asymptotically planar AdS spacetimes is characterized by its ADM mass and tensions.
We define an additional ADM charge $Q$ associated with the scaling Killing vector of AdS, show that $Q$ is given by a certain sum over the ADM mass and tensions and  that $Q$ vanishes on solutions to the Einstein equation with $\Lambda<0$.  The sum rule for the mass and tensions thus established corresponds in an AdS/CFT context to the vanishing of the trace of the boundary stress tensor.
We also show that an 
analogous sum rule holds for local planar sources of stress-energy sources in AdS.  In a simple model consisting of
a static, plane symmetric source
we find that the perturbative stress-energy tensor must be tracefree. 
  \vfill
\vskip 2.mm
\end{quote}
\end{titlepage}

\section{Introduction}
Planar black holes and solitons are important in the phenomenology of the AdS/CFT correspondence \cite{Horowitz:1998ha}.
In addition to the ADM mass $\mass$ such spacetimes are characterized by $D-2$  gravitational tensions $\tension_k$ \cite{Traschen:2001pb,Townsend:2001rg,Harmark:2004ch} in the planar spatial directions $x^k$, which we take to be periodically identified according to 
$x^k\equiv x^k +L$.  In an earlier paper \cite{El-Menoufi:2013pza} establishing Smarr relations for such spacetimes, we observed that the ADM charges satisfy the relation
\begin{equation}\label{zerosum}
\mass  + L \sum_{k=1}^{D-2} \tension_k = 0
\end{equation}
and hence are not all independent.  This property was not connected in \cite{El-Menoufi:2013pza} to any fundamental physical or geometric consideration, but rather followed from the field equations and the detailed expressions for the charges in terms of fall-off coefficients.  The purpose of this paper is to remedy this situation and show how equation (\ref{zerosum}) follows from a scaling symmetry of the asymptotic AdS background.  In the context of AdS/CFT,  the relation (\ref{zerosum}) is equivalent to tracelessness of the boundary stress tensor \cite{Balasubramanian:1999re,Myers:1999psa} and is thus of fundamental significance.  Our goal here is to establish this result using familiar techniques of general relativity in the bulk.

\begin{figure}[!ht]
    \centering
    \begin{floatrow}
      \ffigbox[\FBwidth]{\caption{Spatial slice for Hamiltonian decomposition in proof of first law for asymptotically flat black holes.  The dotted circle represents the outer boundary of the Gaussian surface at infinity.  The inner boundary is the black hole horizon.}\label{bh-fig}}{%
       \includegraphics[width=7.5cm]{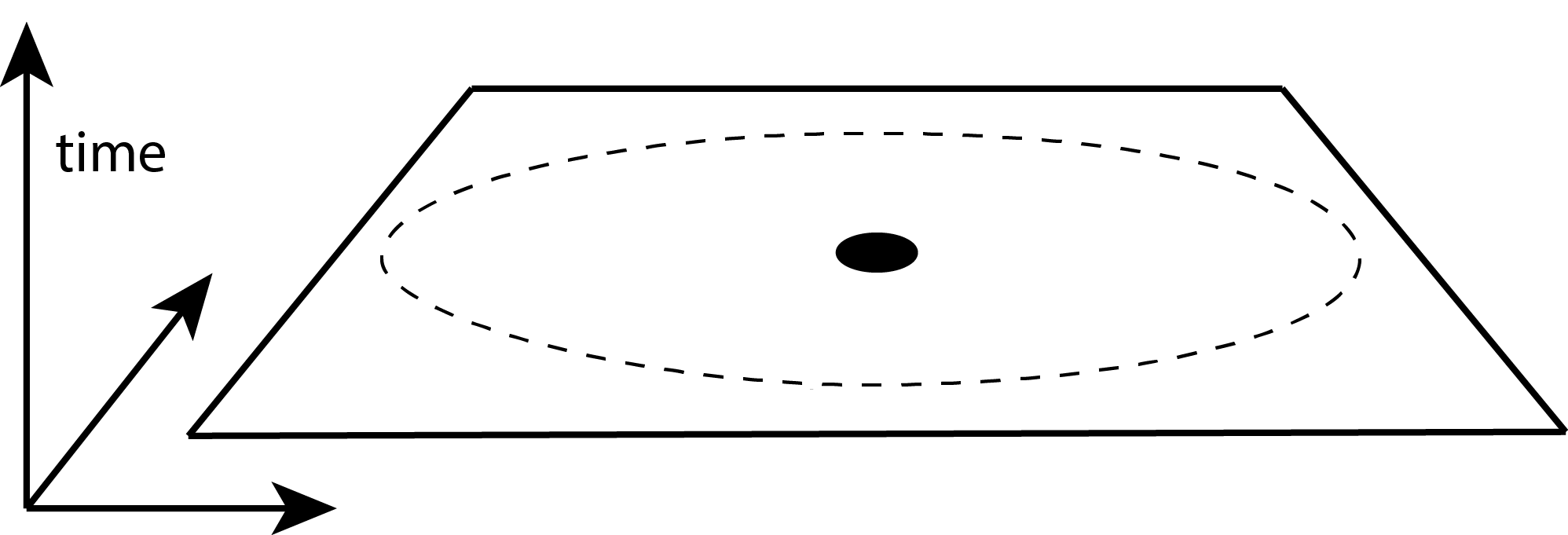}   
      }\qquad
      \ffigbox[\FBwidth]{\caption{Lorentzian slice near spatial infinity used for Hamiltonian decomposition in proof of ADM mass and tension sum rule for planar AdS black holes and solitons.  The dotted box shows the boundary used in the Gauss' law construction.}\label{planar-bh-fig}}{%
     \includegraphics[width=7.5cm]{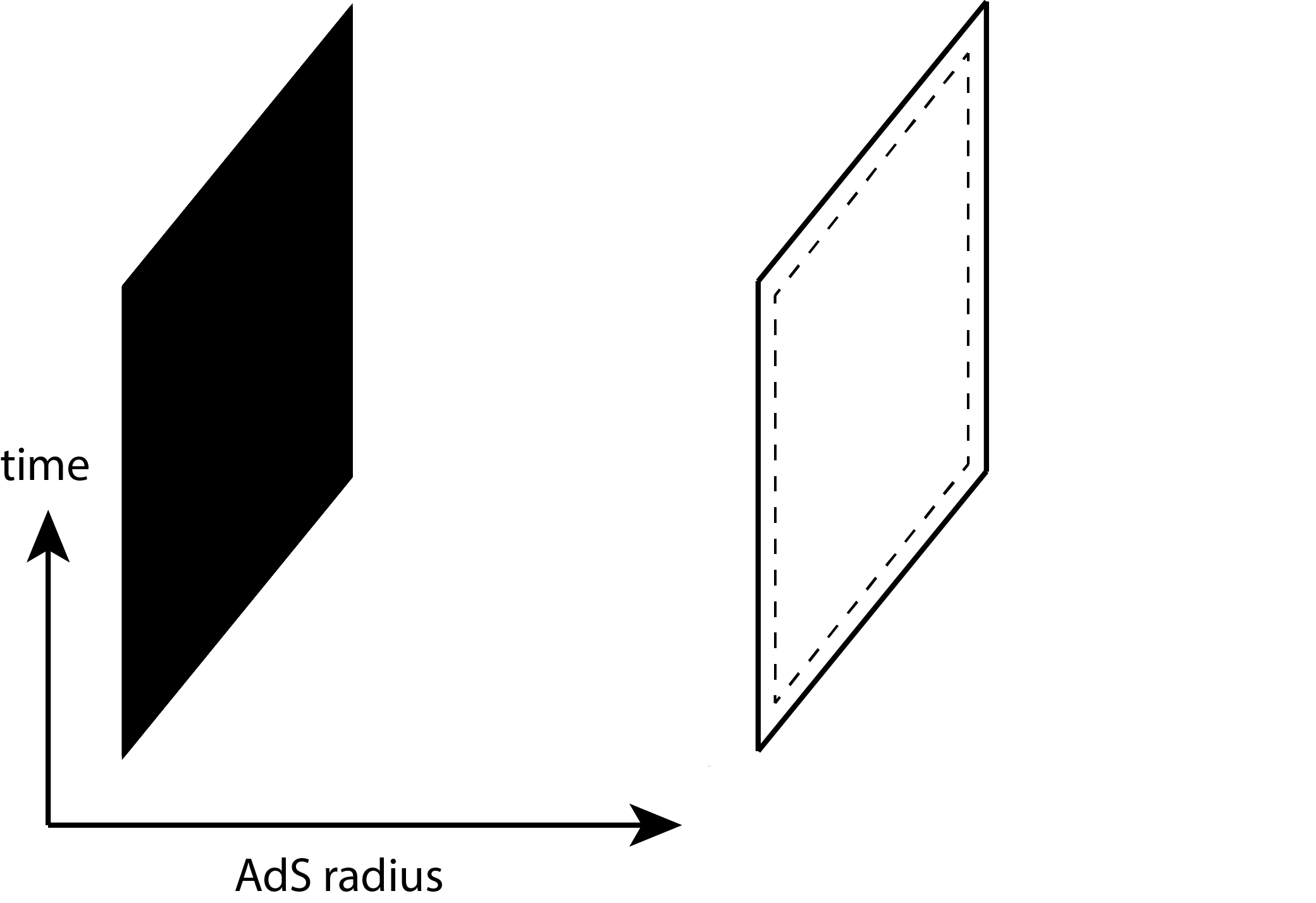}   
      }
    \end{floatrow}
  \end{figure}

To prove the statement (\ref{zerosum}) and connect it with a symmetry of the AdS background, we will make novel use of a Hamiltonian perturbation theory method used in proofs of the first law of black hole thermodynamics \cite{Traschen:1984bp,Sudarsky:1992ty,Traschen:2001pb}.  These constructions are based on a Gauss' law relation that holds for perturbations about  a solution to Einstein's equations when the unperturbed solution has a Killing vector $\xi^a$.  
In the first law context, one considers the Gauss' law relation
on a spacelike hypersurface that is associated with the time translation Killing vector.   A boundary term at infinity, which is expressed in terms of the variations in the ADM charges, is equated with a boundary term at the horizon that is proportional to the variation in the horizon area, resulting in the first law.  This is illustrated in figure (\ref{bh-fig}) where the small black oval representing the black hole horizon and the dotted circle representing a sphere at spatial infinity together form a Gaussian surface on a spacelike slice.

For our application, we will turn this construction on its side,
making use of a Gauss' law relation on a timelike hypersurface that is located entirely in the asymptotic region.  
Specifically, we consider the slicing of AdS by Lorentzian planes and work with a Gauss' law relation coming from the AdS Killing vector that combines translation in the radial direction with a scaling transformation on these planes. 
On a  Lorentzian plane at large AdS radius, this Gauss' law relation implies that a certain boundary integral, which may be expressed as a combination of ADM charges, must vanish.  This yields equation (\ref{zerosum}).  Recalling that tension is minus the pressure, the sum rule can be considered to be a Lorentzian tracelessness condition for the ADM charges associated with this asymptotic scaling Killing vector.  This setup is illustrated in figure (\ref{planar-bh-fig}), where the black surface represents a planar black hole, the white surface represents a Lorentzian slice in the asymptotic regime and the dotted perimeter represents the box-like choice of the Gaussian surface used below.

We also ask what the implications of the constraint on the ADM charges (\ref{zerosum}) are for sources of stress-energy in AdS.  
There is no such constraint relating the mass and tensions in asymptotically flat spacetimes, and planar matter sources in flat spacetime may correspondingly be configured to give independent mass and tensions.
For a localized matter configuration in AdS the environment is approximately flat, and 
one would expect to be similarly able to configure  localized sources.
We will show, however, that planar stress-energy
sources in AdS are in fact constrained by a local version of (\ref{zerosum}).  Hence the ADM mass and tensions of
an asymptotically AdS spacetime can not be independently specified, and the
same is true of sources.

The paper is laid out as follows.  In section (\ref{sumrulesec}) we describe the Hamiltonian perturbation theory construction, present the formulas for the ADM mass and tensions of asymptotically planar AdS spacetimes, and then give the particular application of the construction that establishes the sum rule (\ref{zerosum}).  In section (\ref{sourcesec}), we look at the constraints on planar sources of stress-energy by solving the perturbative field equations.  We also examine the relations between the energy densities and pressures that give rise to the same far field limits as the planar AdS black hole and AdS solution solutions.  In section (\ref{conclude}) we offer some brief conclusions and ideas for future work.

\section{Sum rule for mass and tension in AdS}\label{sumrulesec}

In this section we will introduce and apply Hamiltonian perturbation theory techniques to prove the sum rule (\ref{zerosum}) for the ADM mass and gravitational tensions in asymptotically planar AdS solutions to the Einstein equations with $\Lambda<0$.
ADM gravitational charges are defined in terms of  the asymptotic symmetries of 
a spacetime \cite{Arnowitt:1962hi,Abbott:1981ff}.  We take the asymptotic form of the metric at large AdS radius $r$ to be 
\begin{equation}\label{adsfalloff}
 ds^2 \simeq {r^2 \over l^2 }\left( \eta_{\alpha\beta} + 
  {c_{\alpha\beta} \over r^{D-1}} \right)dx^\alpha dx^\beta   +
 {l^2 \over r^2}\left (1 + {c_r \over r^{D-1} }\right ) dr^2  
 \end{equation}
where $c_{\alpha\beta}$ and $c_r$ are constants.  The AdS background metric in planar slicing is obtained by setting these constants to zero.  Here and below, Latin indices $a,b$ from the beginning of the alphabet will run over all coordinates, while Greek indices 
$\alpha,\beta =0,\dots, D-1$ label coordinates on the constant $r$ planes and Latin indices $i,j= 1,\dots, D-2$ from the middle of the alphabet designate spatial coordinates on the planes.
The asymptotic form of the metric is chosen, as shown below, to yield finite ADM charges. Analytic solutions with these asymptotics include the planar AdS black hole and AdS soliton discussed in \cite{Horowitz:1998ha}.  For simplicity in the following, we will assume that the matrix of fall-off coefficients $c_{\alpha\beta}$ is diagonal.

 \subsection{Hamiltonian perturbation theory}\label{hamsec}
As stated above, the sum rule (\ref{zerosum}) for the ADM mass and gravitational tensions will follow from a new construction making use of the Hamiltonian perturbation theory methods used to establish the first law for stationary black holes \cite{Traschen:1984bp,Sudarsky:1992ty,Traschen:2001pb}.  We briefly recount these methods following the specific treatment in \cite{Traschen:2001pb} which allows for timelike hypersurfaces.
Let ${\cal S}$ denote a family of 
hypersurfaces with normals $n_a$ and decompose the spacetime metric as
 \begin{equation}\label{metricsplit}
g_{ab} = (n\cdot n) n_a n_b +s_{ab}
\end{equation}
where $s_{ab}$ is the induced metric on the hypersurfaces which satisfies $ n^c s_{cb} =0$.  The hypersurfaces ${\cal S}$ may be either spacelike or timelike depending on whether $n\cdot n=-1$ or $+1$.  
The conjugate momentum 
$\pi ^{ab}$ is
related to the  extrinsic curvature of the slice $K_{ab} = s_a{} ^c \nabla _c n_b$ by 
$\pi^{ab} =(n\cdot n)\sqrt{|s|}(K s^{ab} -K^{ab})$.

The differences between the hypersurface metric and 
momentum $(s_{ab},\pi^{ab})$ and the corresponding background AdS quantities denoted $(\bar s_{ab},\bar\pi^{ab})$ are given by $h_{ab} =s_{ab} -{\bar s}_{ab}$ and 
$ p^{ab}=\pi ^{ab} - {\bar \pi}^{ab}$. 
Let $\xi^a$ be  a Killing vector of the background AdS metric $\bar g_{ab}$ and hence is an asymptotic symmetry of
$g_{ab}$.  It can be decomposed into  components normal and  tangential to $\cal S$ as
\begin{equation}\label{kvdecomp}
\xi ^a = Fn^a +\beta ^a .
\end{equation}
The ADM gravitational charge
associated with this symmetry is defined as an integral over the $(D-2)$-dimensional
 boundary of ${\cal S}$ in the asymptotic region
\begin{equation}\label{chargedef}
Q = -{1\over 16 \pi}\int  _{\partial{\cal S}_\infty} da_c B^c 
\end{equation}
where  the boundary vector $B^a$ is given by 
 \begin{equation}\label{boundaryvec}
B^a=F({D}^ah-{D}_bh^{ab})-h{D}^aF+h^{ab}{D}_bF+
{1 \over \sqrt{{\bar s}}}\beta^b({\bar \pi}^{cd}h_{cd}{\bar s}^a{}_b-2{\bar \pi}^{ac}h_{bc}-2p^a{}_b)
\end{equation}
and $D_a$ is the covariant derivative operator with respect to the background metric $\bar s_{ab}$ on the slice $\cal S$.
Note that this boundary vector depends on both the choice of Killing vector and the choice of slice. 

We will also need the following related result. Let $\gzero$ be an asymptotically planar AdS spacetime
as in (\ref{adsfalloff}) that solves the Einstein equations with $\Lambda<0$ and has the Killing vector $\xi^a$.   
For example, $\gzero$ could be the planar AdS black hole  and $\xi^a$ the time translation Killing field.  
Assume that the metric $g_{ab}$ above is perturbatively close to $\hat g_{ab}$ and also solves the equations of motion.
One can then show that a Gauss law relation  \cite{Traschen:1984bp,Sudarsky:1992ty,Traschen:2001pb} given by
\begin{equation}\label{gauss}
\int _{\partial\Sigma} da_c B^c =0.
\end{equation}
holds, where $\partial\Sigma$ is the boundary, possibly having multiple components, of a connected region $\Sigma$  contained within the slice ${\cal S}$.  The boundary vector $B^c$ is as given in (\ref{boundaryvec}) with the background fields ${\bar s}_{ab}$
and ${\bar \pi}^{ab}$ replaced by the quantities ${\hat s}_{ab}$ and ${\hat \pi}^{ab}$ respectively and likewise
$h_{ab}=s_{ab}-{\hat s}_{ab}$ and $p^{ab}=\pi^{ab}-{\hat\pi}^{ab}$
now representing the perturbations away from
the background $\gzero$.  

The Gauss law relation (\ref{gauss}) has been used to prove the first law \cite{Sudarsky:1992ty} by taking $\hat g_{ab}$ to be the metric a stationary black hole solution and the volume $\Sigma$ to extend from the black hole horizon out to infinity.  The boundary term at infinity gives the variation in the ADM mass, while the boundary term at the horizon is proportional to the surface gravity times the variation in the horizon area.  We will make use of (\ref{gauss}) in a different way below to prove the sum rule (\ref{zerosum}) for the ADM mass and tensions.

\subsection{ADM mass and tension}
 The general formula for an ADM charge given in (\ref{chargedef}) depends on a choice of asymptotic symmetry and also on a choice of hypersurface, with
 different choices yielding different charges.  The ADM mass $\mass$ is obtained 
by choosing a constant time slice together with the asymptotic time translation symmetry.
Similarly, the ADM tension $\tension_k$ \cite{Traschen:2001pb,Townsend:2001rg,Harmark:2004ch} is obtained by taking a slice of constant spatial coordinate $x^k$ and the asymptotic translational symmetry in the $x^k$ direction.
The boundary integral for $\mass$ involves integrating over the $D-2$ spatial directions in
the plane.
If these have infinite extent then, although the mass per unit volume is finite, the total ADM mass is infinite.  In order to avoid this, we will take the spatial directions along the plane to be compact with periodic boundary conditions and coordinate ranges
$-L/2 \leq x^k < L/2$.  
The definition of  $\tension _k$ involves an integration over the
coordinates $x^l$ with $l\neq k$ and additionally an integration over the time direction.   
The integrand in (\ref{chargedef}) is time independent and the time integration may therefore be suppressed to define $\tension_k$ as a tension per unit time.

Substituting the asymptotic form of the metric (\ref{adsfalloff}) into the expression (\ref{chargedef}) for the ADM charge  using these choices for the slice and the asymptotic symmetry gives
\begin{eqnarray}\label{masstension}
\mass & = &{L^{D-2} \over 16\pi l^D}\left[ (D-2) c_{r} +(D-1)  \sum_j c_{jj}
\right]  \\  
\tension _k & = &{L^{D-3} \over 16\pi  l^D}\left[ (D-2) c_{r} +(D-1) ( -c_{tt} +\sum_{j\neq k}c_{jj}
  ) \right] 
\end{eqnarray}
For the purposes of the next section, we note that 
\begin{eqnarray}\label{chargesum}
 \mass +L\sum _{k=1}^{D-2} \tension _k & = & (D-1)(D-2) {L^{D-2} \over 16\pi l^D} \left( c_r +c_\alpha{}^\alpha\right)\\
&=&  (D-1)(D-2) {L^{D-2} \over 16\pi l^D}\left( \lim_{r\rightarrow\infty}r^{D-1} \gamma\right) \nonumber
\end{eqnarray}
where $\gamma_{ab} \equiv g_{ab} - {\bar g}_{ab}$ is the difference between the spacetime metric and the AdS background  in the asymptotic region and $\gamma=\bar g^{ab}\gamma_{ab}$ is the trace of this quantity.
 
 This is an interesting result, because 
when $g _{ab}$ is a solution to the field equations, which means that $\gamma_{ab}$ is a 
solution to the linearized equations  in the far field,  the trace $\gamma$ must vanish
\cite{El-Menoufi:2013pza}. Therefore, equation (\ref{chargesum}) implies that the trace of the ADM charges is zero for solutions  to the field equations, which is equation (\ref{zerosum}).

\subsection{AdS sum rule}
We are now ready to present a construction that provides a geometrical understanding
for the sum rule (\ref{zerosum}) satisfied by the ADM charges. This is 
 a non-standard first law construction that results from choosing the Killing vector in (\ref{gauss}) 
 to be the scaling Killing vector $\xi^a$ of AdS
\begin{equation}\label{kv}
\xi^a= x^\alpha {\partial \over \partial x^\alpha } -r {\partial \over \partial r}
\end{equation}
  and ${\cal S} $ to be
a plane of constant $r$  with unit spacelike normal in the radial direction. 
These choices are non-standard because ${\cal S}$ does not intersect the boundary
 at infinity. However by taking ${\cal S}$ to lie in the asymptotic region at large AdS radius $r$,
we will get information about the standard ADM charges.

Following the conventions of the last section, we then have $n_r=l/r$ and $\bar s_{\alpha\beta}= (r^2 /l^2 ) \eta _{\alpha\beta} $ which gives $\bar{K}_{\alpha\beta}= \bar s_{\alpha\beta} /l$ and
$\bar\pi^{\alpha\beta} = (D-2)\sqrt{-\bar s}\, \bar s^{\alpha\beta}$.
Decomposing the Killing vector $\xi^a$ according to (\ref{kvdecomp}) gives $F= -l$ and  $\beta^\alpha = x^\alpha$.
Substituting  these quantities into (\ref{boundaryvec}) for the boundary vector $B^a$ gives
 \begin{equation} \label{bforv}
B^\alpha  =  -l (D^\alpha h - D_\beta h^{\beta\alpha} )
+  \left( {D-2\over l} ( h \bar s^\alpha {}_\beta -2 h^\alpha {}_\beta )
-{2\over\sqrt{-\bar s}}p^\alpha{} _\beta \right) x^\beta
\end{equation}
We take the slice $\cal S$ to be at sufficiently large radial coordinate $r=r_0$ such that the spacetime metric is given accurately by  
equation (\ref{adsfalloff}). The perturbation to the spacetime metric is then found to be given in terms of the Hamiltonian variables by 
\begin{eqnarray}
h_{\alpha\beta}&=&{c_{\alpha\beta}\over l^2r^{D-3}}\\
p^{\alpha\beta}&=&-{\sqrt{-\bar s}\, l\over r^{D+1}}\left( {D-3\over 2}c^{\alpha\beta}+\left[(D-1)c_r +{D-2\over 2}c_\rho {}^\rho\right]\eta^{\alpha\beta}\right)
\end{eqnarray}
Because the metric perturbation $h_{\alpha\beta}$ is constant in the directions along the plane $r=r_0$ and the background metric $\bar s_{\alpha\beta}$ is flat, the terms inside the first set of parenthesis in (\ref{bforv}) vanish identically.  The remaining terms then yield the expression
\begin{equation}
B^\alpha = {1\over lr^{D-1}}\left( (D-2)c_r\delta^\alpha_\beta +(D-1)c_\rho{}^\rho\delta^\alpha_\beta - (D-1)c^\alpha{}_\beta\right)\, x^\beta
\end{equation}

For this construction, we take the volume of integration $\Sigma$ implicit in the Gauss law relation (\ref{gauss}) to be the Lorentzian box  $-x_0 ^\nu \leq x^\nu \leq x_0 ^\nu$.  Here, each $x_0^\mu<L/2$ such that the box is well-defined given the periodic identifications\footnote{Note that the 
symmetry generated by the Killing vector $\xi^a$ is broken by the choice of periodic boundary conditions on the planar slices.
However, the construction relies only on the existence of the Killing vector within a local region.} of the slice ${\cal S}$.
The $D-2$ dimensional boundary $\partial\Sigma$ is then the union of $2(D-1)$ planar `box tops' located at $\pm\, x_0^\nu$ for $\nu=0,1,\dots,D-2$.
The normal to the boundary component at $\pm\, x_0^\alpha$ is given by $m^{(\alpha,\pm)}_\beta=\pm (r_0/l)\delta^\alpha_\beta$, and one finds on this portion of $\partial\Sigma$ that
\begin{equation}\label{boundary}
m^{(\alpha,\pm)}_\beta B^\beta =\pm\, {1\over l^2 r_0^{D-2}}\left( \left[ (D-1)\sum_{\beta\neq\alpha}c_\beta{}^\beta +(D-2)c_r\right]\, (\pm\, x_0^\alpha)
-(D-1)\sum_{\beta\neq\alpha}c^\alpha{}_\beta\, x^\beta\right)
\end{equation}
Because the region of integration along the boundary is even in the remaining coordinates $x^\beta$ with $\beta\neq\alpha$, while the last term in (\ref{boundary}) is odd, this term integrates to zero.  For the remaining terms, the changing sign of the normal between opposite boundaries goes together with the changing sign of the explicit factor of $\pm\, x_0^\alpha$ in the integrand to produce the overall result for $I=\int _{\partial\Sigma}da_c B^c$,
\begin{equation}\label{final}
I = {V_0\over l^D}(D-1)(D-2)\left(c_r +c_\alpha{}^\alpha\right)
\end{equation}
where $V_0$ is the Lorentzian volume $\prod_{\mu=0}^{D-2}(2x_0^\mu)$.  Since the Gauss law relation (\ref{gauss}) implies that the integral $I$ must vanish, and since equation (\ref{chargesum}) relates  the sum of the ADM mass and tensions  to the same combination of falloff coefficients appearing in $I$, we have proved the sum rule 
\begin{equation}
 \mass +L\sum _{k=1}^{D-2} \tension _k =0.
\end{equation}
In summary, we have seen that this result follows generally from a first-law type construction based on the AdS scaling Killing vector (\ref{kv}).  
In fact, the quantity $Q=-{1\over 16\pi} I$ is the ADM charge for the scaling Killing vector, defined with respect to the Lorenztian hypersurfaces at large AdS radius.
We have shown here that $Q$ can be expressed as a sum of other ADM charges and that this sum necessarily vanishes as a consequence of the equations of motion.

 \section{Perturbative sources in AdS}\label{sourcesec}
 In this section, we show that  a perturbative source of stress-energy in AdS
 must satisfy a local version of the sum rule (\ref{zerosum}). 
 We consider a  static, planar source of stress-energy described by a radially dependent energy density
 $\delta \rho (r)$ and pressures $\delta p_i (r)$.  We assume that the sources are localized, so that 
 metric has the form (\ref{adsfalloff}) in the asymptotic region.  
 
 If instead, we were considering such planar, localized sources with $\Lambda=0$, then
the density and pressures 
would be independently specifiable quantities, corresponding to the independence of the ADM mass $\mass$ and  tensions $\tension _k$.
Examples with  $1+1$ and $2+1$ dimensional planar sources were studied respectively in references \cite{Harmark:2003eg}
and \cite{Kastor:2006ti}. On the other hand, in AdS we have seen that the ADM mass and tensions are not independent.
One might be concerned that this global property could conflict with an ability to freely specify local sources
in AdS.  We will show in this section that this is not the case, that local
sources are, in fact, constrained in an analogous way to the gravitational charges.
 
\subsection{Solving the linearized field equations with planar sources}\label{linsol}
We take the sources to be  localized around $r=0$ and the metric to have the form
 \begin{equation}\label{adspert}
 ds^2 = \left({r^2 \over l^2 } \eta_{\alpha\beta} + \gamma_{\alpha\beta} (r) \right)dx^\alpha dx^\beta  
 +\left({l^2 \over r^2}+  \gamma _{rr} (r)\right)dr^2 
\end{equation}
with  $\gamma_{ab}\ll 1$ and $\gamma _{\alpha\beta}$ assumed to be diagonal for simplicity.
One can simply obtain $(D-1)$ of the non-trivial, diagonal components of the Einstein equations for (\ref{adspert}) linearized around the AdS background using the Hamiltonian perturbation theory formalism of section (\ref{hamsec}).
Equation (\ref{gauss}) is now modified as in \cite{Traschen:2001pb} to include
the perturbative stress energy source, giving
\begin{equation}\label{sourcegauss}
D_a B^a =  16\pi (n\cdot n) F\, \delta T_{ab}\,  n^a n^b ~.
\end{equation}
Here, the non-zero components of the perturbative stress energy tensor are $\delta T_t{}^t=-\delta\rho$ and $\delta T_k{}^k=\delta p_k$. We use
a simple model for the radial dependence of the sources that has a $\delta$-function limit.
  It is straightforward to check that in this limit conservation of
  stress-energy 
and vanishing of  the radial
pressure at the outer boundary of the source force
the radial component of the pressure to be zero.   This is why we have set $\delta p_r=0$ from the start.
Equation (\ref{gauss}) can now be used successively on surfaces of
constant  planar coordinate $x^\alpha$, with $\alpha =0,1,\dots ,D-2$, to obtain a set of $(D-1)$ Poisson-type equations. 
The system of perturbation equations is then completed by using the linearized trace of the Einstein equation.
In the following, we will 
denote particular components of the metric perturbation as $\gamma _k{}^k$
with no sum implied on the index $k$, while
 $\gamma =\sum _{a=0}^{D-1} \gamma _a{}^a$ will denote the trace of the metric perturbation.
  
  First, consider a  surface of constant planar spatial coordinate
  $x^k =x_0^k$ and take
  the Killing vector in the construction of the boundary vector $B^a$ to be 
  ${\xi} =  {\partial / \partial x^k} $. The decomposition (\ref{kvdecomp}) of $\xi$
  with respect to this surface is then given by $F=r/l$ and $\beta ^a =0$. 
  One further has $h_{ab}= \gamma _{ab}$ for $a,b \neq k$, while $h_{kk}\equiv 0$. 
 One then finds that equation (\ref{gauss}) then becomes
 \begin{equation}\label{poissonx}
 {1\over  r^{D-2}}{\partial \over \partial r}\left( r^D {\partial \over \partial r }\left( \gtt + 
 \sum _{i \neq k}\gii \right) 
 -(D-2)  r^{D-1} \grr \right) =16\pi l^2 \delta p_k
 \end{equation}
 and we may successively consider $k=1,\dots, D-2$.
If we now carry out the same construction for a surface of constant time $t= t_0$, with the Killing vector $\xi=\partial/\partial t$, 
then equation (\ref{gauss}) becomes
 \begin{equation}\label{poissont}
 {1\over r^{D-2}}{\partial \over \partial r}\left( r^D {\partial \over \partial r } \sum_i \gii 
  - (D-2) r^{D-1} \grr \right) = -16\pi l^2\delta \rho ~.
 \end{equation}
Together equations (\ref{poissonx}) and (\ref{poissont}) constitute $(D-1)$ perturbation equations for the diagonal matrix $\gamma_{\alpha\beta}$ and $\gamma_{rr}$.
To fill out a complete set of $D$ independent equations, we also consider the
linearized trace of  the Einstein equation $\nabla _a \nabla ^a \gamma
-\nabla _a \nabla _b \gamma ^{ab} +2 \Lambda\gamma /(D-2)  =\s $,
where $\s$ is defined by $\s ={16\pi l^2\over D-2} ( -\delta \rho +\sum _k \delta p_k )$.
 One finds that this equation is given explicitly by
\begin{equation}\label{traceeinstein}
 {1\over  r^{D-2}}{\partial \over \partial r}\left( r^D {\partial \over \partial r }( \gamma  -\gamma_r{}^r )
+r^{D-1}(\gamma -D\gamma _r{} ^r )   \right)  -(D-1) \gamma  =\s ~.
 \end{equation}

We see that  equation (\ref{traceeinstein}) only involves the quantities $\gamma$ and $\gamma _r^r$. A second
  linear combination that only involves $\gamma$ and $\gamma _r^r$ can be obtained by  adding equation
  (\ref{poissont}) to the sum of equations (\ref{poissonx}) over all $k$, which gives
\begin{equation}\label{partialtrace}
 {1\over  r^{D-2}}{\partial \over \partial r}
 \left( r^D {\partial \over \partial r }( \gamma  -\gamma_r^r )
-(D-1) r^{D-1}\gamma _r ^r )  \right)  = \s ~.
 \end{equation}
Note that while equations (\ref{traceeinstein}) and (\ref{partialtrace}) 
represent  different  linear combinations of components of the Einstein equation, their right hand sides
 both contain the same combination of sources $\s$.  This happens because the radial pressure
 vanishes.
 
In order to solve the perturbation equations,  it is convenient to work with equations (\ref{traceeinstein}) and (\ref{partialtrace}), together with
 the $D-2$ equations obtained by taking the differences of equations (\ref{poissonx}) and (\ref{poissont})
 for each value of $k$, 
 which are given by
 \begin{equation}\label{poissontx}
 {1\over r^{D-2}}{\partial \over \partial r}\left( r^D {\partial \over \partial r }
 (\gamma _t{}^t   - \gamma _k{}^k ) \right) = 16\pi l^2(\delta \rho +\delta p_k)
 \end{equation}
 with $k=1,\dots, D-2$.
 We will solve this system of equations by first determining the solutions outside of the source region,
 then finding the interior solutions and finally matching the interior and exterior solutions.
 
In the source-free region the solution to  equation (\ref{poissontx})  for each $k$ is
$ (\gamma _t{}^t   - \gamma _k {}^k ) =a_k r^{-D+1} + \alpha_k$, where $a_k$  and $\alpha_k$
are constants. The additive constant $\alpha_k$ can be set to zero by rescaling the
coordinates, so we will not include it.
It follows  that 
 \begin{equation}\label{homsol}
 \gamma _t {}^t = - {c_t \over r^{D-1}  } +\psi (r) \  , \quad  \gamma _k {} ^k=  {c_k \over r^{D-1} } +\psi (r) 
 \end{equation}
 for $k=1,...,D-2$,
 where $\psi (r)$ is an undetermined function
 and $c_t$ and  $c_k $ are constants. 
 
Inspection of equations (\ref{traceeinstein}) and (\ref{partialtrace})
shows that they are a coupled system of one second order and one first order differential equation for the
functions $ \gamma -\gamma _r{}^r $ and $ \gamma _r{}^r$. Moreover, the 
system collapses in an interesting way. The difference between the two equations is
\begin{equation}\label{diff}
\Psi \equiv \left[  r \partial_r ( \gamma  -\gamma _r{}^r ) -(D-1)\gamma _r{}^r  \right] = 0 
\end{equation}
Hence in both the interior and exterior regions $\gamma _r{}^r$ is determined by
the other components to be 
\begin{equation}\label{grr}
\gamma _r{}^r   ={r \partial_r ( \gamma _t{}^t  + \sum  \gamma _i{} ^i ) \over D-1} 
\end{equation}
Further, equation (\ref{partialtrace}) can be rewritten as
\begin{equation}\label{rewrite}
 r\partial_r \Psi +(D-1) \Psi =\s
\end{equation}
Consistency between equations (\ref{diff}) and (\ref{rewrite}) in the interior region
then requires that the sources must satisfy $\s  =0$. 
We have therefore found that in order to have a perturbative solution to the equations of motion of the assumed form the
sources must satisfy
\begin{equation}\label{sourcezero}
 \delta\rho  - \sum_{k=1}^{D-2}  \delta p_k = 0,
\end{equation}
which is a local version of the sum rule (\ref{zerosum}) satisfied by the  ADM charges. 
This constraint is forced by the field equations, and is distinct from the situation
for planar sources with $\Lambda =0$. 
It resolves any possible conflict between allowed local sources and the global properties of asymptotically planar AdS spacetimes.

Although demonstrating the existence of the constraint (\ref{sourcezero}) on sources is the main result of this section, we go on to complete the solutions outside the source region and the matching of interior to exterior solutions.  Equation (\ref{grr}) tells us that 
the function $\psi$
in (\ref{homsol}) makes a contribution of $r\partial _r \psi$ to the metric function $\gamma _r{}^r$.  The function
$\psi$ is undetermined by the equations of motion and represents a pure gauge degree of freedom. Indeed,
it is straightforward to check that this mode can be set equal to zero by
the coordinate transformation $x^{a \prime} = x^a +W^a$ with $W_r = -(l^2/2r)\psi$
 and $W_\alpha =0$. Hence, the exterior solution is given by
\begin{equation}\label{out}
  \gamma _t{}^t = - {c_t\over r^{D-1}} \  , \quad
   \gamma _k{}^k = {c_k\over r^{D-1}} \  , \quad
 \gamma _r{}^r = {c_r \over r^{D-1}}  
\end{equation}
with $c_r = c_t -\sum_i c_i $.  The function $\psi$ may also be fixed by choosing to work in {\it e.g.} harmonic gauge.

To fix the interior solution, we use a simple model for planar sources.   Following the analyses  \cite{Harmark:2003eg,Kastor:2006ti} for $\Lambda=0$,
we assume that the density 
and tangential pressures 
are constant  out to a matching radius $r=\epsilon$ and vanish outside this radius.  We then take a limit of $\epsilon$ tending to zero, with the volume integrals of the sources held fixed.
In this limit, the sources become $\delta$-functions of fixed magnitude.
While a particle-like $\delta$-function source is characterized solely by its mass,
a planar $\delta$-function source has tangential pressures (or equivalently tensions) as well. 
Explicitly, we take sources such that
\begin{equation}\label{massnorm}
\delta \rho  = {m\over V_\epsilon} \  , \quad \delta p_k   = {P _k\over V_\epsilon}
\end{equation}
where $V_\epsilon =   \epsilon ^{D-1}L^{D-2}/ (D-1) l^{D-2}$ and 
 $m$ and $P_k$ are the fixed total mass integrated pressures of the planar source. 

In the interior region the general solution to equation (\ref{poissontx}) with these sources is then given by
\begin{equation}\label{insoltx}
   \gamma_t{} ^t -\gamma_k{} ^k  = 
  {  16\pi l^2 \over D-1}  (\delta\rho + \delta p_k ) \ln (r / \epsilon ) + \beta_k 
\end{equation}
 where the $\beta _k$ are  constants. 
 Continuity of the functions
  $  \gamma_t{} ^t -\gamma_k{} ^k $  and their  first derivatives at $r=\epsilon$ determines
  these constants to be given by $\beta _k = -16\pi l^2 (\delta\rho + \delta p_k )/(D-1)^2$ and also that 
\begin{equation}\label{ctck}
 c_t + c_k   = {16\pi l^D \over (D-1)L^{D-2} }
 (m +P_k ) .
 \end{equation}
To determine the coefficients $c_t$ and $c_k$
individually, the residual gauge freedom discussed above must be fixed.
We choose a gauge such that $c_r =0$, which is similar to synchronous
gauge in cosmology. In order to accomplish this, let $x^{a \prime} = x^a +V^a$ with $V_r =Kr^{-D} $
 and $V_\alpha =0$.   Taking $K= - l^2 c_{r}/2(D-1) $ then gives $c_{r}^\prime=0$
 and shifts the other coefficients by $c_{\alpha}^\prime =c_{\alpha} +c_{r} /(D-1)$.
This amounts to setting $c_r =0$ and relabeling the other coefficients. 
Given this, the fact that $\gamma =0$ in the exterior (see equation (\ref{out})) implies that
 \begin{equation}\label{csum}
 -c_t + \sum_i c_i =0
 \end{equation}
 Summing equation (\ref{ctck}) over $k$ and using (\ref{csum}) and (\ref{sourcezero}) then gives
\begin{equation}\label{coeffsmatch}
c_k = {16\pi l^D\, P_k  \over (D-1) L^{(D-2)}} ,\qquad
c_t = {16\pi l^D \,  m \over (D-1) L^{(D-2)}}
\end{equation}
The interior solution (\ref{insoltx}) can be processed similarly to fully specify the 
metric components $\gamma_\alpha {}^\beta$ independently given the gauge choice $\gamma_r {} ^r=0$.

This completes the linearized solution with static planar sources. To summarize, we have found
that the equation of state of perturbative static sources in AdS can not be arbitrarily specified. 
The sources must satisfy the zero trace condition (\ref{sourcezero}), which is the local version
of the sum rule (\ref{zerosum}) satisfied by the ADM charges. 

\subsection{Connecting planar black holes and solitons to sources}
In this section, we ask what perturbative sources produce the same far field limit
as the planar black hole and AdS soliton  \cite{Horowitz:1998ha}.   
The planar black hole metric is given by
\begin{equation}\label{bhmetric}
 ds^2 = {r^2 \over l^2 }\left( -dt^2 \left( 1- {r_0 ^{D-1} \over r ^{D-1}} \right)  +
   \delta_{ij} dx^i dx^j  \right)+
 {l^2 \over r^2} {dr^2 \over \left( 1- {r_0 ^{D-1} \over r^{D-1}}\right ) }  
 \end{equation}
 where $r_0$ is the horizon radius.  
 The ADM mass and tensions were found in \cite{El-Menoufi:2013pza} to be given by
\begin{equation}\label{bhmt} 
  \mass = {(D-2)L^{D-2}\, r_0^{D-1} \over 16\pi l^D},  \quad  L\ten _i = - {L^{D-2}\, r_0^{D-1}\over 16\pi l^D}
\end{equation}
with $i=1,\dots, D-2$.  One sees that for the planar black hole the tensions are related to the mass according to $L\ten _i=-M/(D-2)$, and it is straightforward to check that the sum rule (\ref{zerosum}) is satisfied.

The AdS soliton metric is obtained from (\ref{bhmetric}) via the double analytic continuation $t \rightarrow iz$ and
$x^{D-2} \rightarrow it$.   The $z$-direction is taken to be compact, $z\equiv z+L_z$, with $L_z=4\pi l^2/(D-1)r_0$ to ensure smoothness at $r=r_0$.
The mass and tensions of the AdS soliton were found \cite{El-Menoufi:2013pza}  to be
\begin{equation}\label{solitonmt} 
  L_z\ten_z = {(D-2)L^{D-3}L_z\, r_0^{D-1} \over 16\pi l^D},
 \quad \mass = L \ten _i = -  {L^{D-3}L_z\, r_0^{D-1} \over 16\pi l^D}
\end{equation}
where now $ i=1,\dots, D-3$.  We see that the mass of the AdS soliton is negative and that $M=L\ten_i= -L_z\ten_z/(D-2)$, so that the sum rule (\ref{zerosum}) is again satisfied.

The sources $\delta \rho$ and $\delta p_k$ that reproduce the far field limits of planar black hole and AdS soliton metrics
are found to have the same relations as the corresponding ADM mass and tensions. The planar black hole metric (\ref{bhmt}) has 
$c_t = c_r =r_0 ^{D-1}$ and $c_k =0$ for  $k=1,\dots, D-2$.
The perturbative solutions in section (\ref{linsol}) have been found using the gauge choice $c_r =0$.   To compare the far field of the planar black hole with our perturbative solutions, it is then necessary to make a coordinate transformation in the far field of (\ref{bhmt}) to reach this gauge. 
This is achieved by taking $x^{a \prime} = x^a +V^a$ with $V_r =-r_0 ^{D-1} /2(D-1) r^D $
and $V_a =0$ otherwise.
One then finds that  
the fall off coefficients in the new coordinates are given by
\begin{equation}\label{newca}
c_r ' =0,\qquad c_t '=  {(D-2)r_0 ^{D-1} \over D-1}  \  , \qquad c_k '= {r_0 ^{D-1} \over D-1} \   , 
\end{equation}
Consulting equation (\ref{coeffsmatch}), we see that the relation between these
coefficients matches that between  the ADM mass and tension for the planar black hole,
using the relation $P_k = - L\ten_k$. Repeating these steps
for the AdS soliton metric (\ref{solitonmt}) similarly shows that local sources in that case are related in the same way as the ADM mass and tensions.

\section{Conclusions}\label{conclude}

We have shown that the ADM mass and tensions of asymptotically planar AdS spacetimes satisfy the sum rule (\ref{zerosum}).  
From an AdS/CFT point of view this is equivalent to the vanishing of the trace of the boundary stress tensor.  Here, however, we have seen how this result arises for the bulk ADM charges as a consequence of the existence of the AdS scaling Killing vector (\ref{kv}).  Our result can, in fact, be interpreted as the vanishing of the ADM charge $Q$ associated with this asymptotic scaling symmetry.  It followed from a novel use of the Hamiltonian perturbation theory techniques of section (\ref{sumrulesec}) applied to a Lorentian slice residing entirely in the asymptotic region.
We have further seen how the linearized field equations forces perturbed sources of stress energy, having the planar symmetry, to satisfy a local version of this sum rule given by (\ref{sourcezero}).  This says that the perturbative stress tensor $\delta T_a{}^b$ must be tracefree.

One interesting direction for further work is to relax the asymptotically AdS boundary
conditions and allow the falloff coefficients in  (\ref{adsfalloff}) to be functions of the coordinates $x^\alpha$ on the Lorentzian planes. 
The construction here shows that the sum $\mass + L\sum _i \ten _i$
will no longer be zero, but will receive a contribution from the geometry on the Lorentzian planes.
This can be seen explicitly from equation (\ref{boundaryvec}).
Such relaxed boundary conditions would also necessitate generalizing the definition of the ADM
 charges, as they would  also now depend on the  $x^\alpha$. A second generalization 
 would be to study black holes with horizons that extend to infinity, such as those studied in references \cite{Figueras:2011va,Haehl:2012tw,Fischetti:2012ps}. In this case, our usual structure for defining {\it e.g.}  the mass of the spacetime does not apply, because the black hole
 horizon extends all the way to infinity. A final area for further work would be to investigate how the 
 constraint on stress-energy sources generalizes when the system is time dependent.

\subsection*{Acknowledgements}

 JT thanks the Kavli Institute 
for Theoretical Physics, where a portion of this work was done, for their hospitality.  This research was supported in part by NSF grants PHY05-55304 and PHY11-25915.


\begin{thebibliography}{99}

\bibitem{Horowitz:1998ha} 
  G.~T.~Horowitz and R.~C.~Myers,
  ``The AdS / CFT correspondence and a new positive energy conjecture for general relativity,''
  Phys.\ Rev.\ D {\bf 59}, 026005 (1998)
  [hep-th/9808079].



\bibitem{Traschen:2001pb}
  J.~H.~Traschen and D.~Fox,
  ``Tension perturbations of black brane spacetimes,''
  Class.\ Quant.\ Grav.\  {\bf 21}, 289 (2004)
  [arXiv:gr-qc/0103106].
  
\bibitem{Townsend:2001rg} 
  P.~K.~Townsend and M.~Zamaklar,
  ``The First law of black brane mechanics,''
  Class.\ Quant.\ Grav.\  {\bf 18}, 5269 (2001)
  [hep-th/0107228].

\bibitem{Harmark:2004ch} 
  T.~Harmark and N.~A.~Obers,
  ``General definition of gravitational tension,''
  JHEP {\bf 0405}, 043 (2004)
  [hep-th/0403103].

\bibitem{El-Menoufi:2013pza} 
  B.~M.~El-Menoufi, B.~Ett, D.~Kastor and J.~Traschen,
  ``Gravitational Tension and Thermodynamics of Planar AdS Spacetimes,''
  arXiv:1302.6980 [hep-th].
  
\bibitem{Balasubramanian:1999re} 
  V.~Balasubramanian and P.~Kraus,
  ``A Stress tensor for Anti-de Sitter gravity,''
  Commun.\ Math.\ Phys.\  {\bf 208}, 413 (1999)
  [hep-th/9902121].

  
\bibitem{Myers:1999psa} 
  R.~C.~Myers,
  ``Stress tensors and Casimir energies in the AdS / CFT correspondence,''
  Phys.\ Rev.\ D {\bf 60}, 046002 (1999)
  [hep-th/9903203].


\bibitem{Traschen:1984bp}
  J.~H.~Traschen,
  ``Constraints On Stress Energy Perturbations In General Relativity,''
  Phys.\ Rev.\  D {\bf 31}, 283 (1985).

  
\bibitem{Sudarsky:1992ty}
  D.~Sudarsky and R.~M.~Wald,
  ``Extrema of mass, stationarity, and staticity, and solutions to the Einstein
  Yang-Mills equations,''
  Phys.\ Rev.\  D {\bf 46}, 1453 (1992).


\bibitem{Arnowitt:1962hi} 
  R.~L.~Arnowitt, S.~Deser and C.~W.~Misner,
  ``The Dynamics of general relativity,''
  gr-qc/0405109.

\bibitem{Abbott:1981ff} 
  L.~F.~Abbott and S.~Deser,
  ``Stability of Gravity with a Cosmological Constant,''
  Nucl.\ Phys.\ B {\bf 195}, 76 (1982).
  
 


  
\bibitem{Harmark:2003eg} 
  T.~Harmark and N.~A.~Obers,
  ``Phase structure of black holes and strings on cylinders,''
  Nucl.\ Phys.\ B {\bf 684}, 183 (2004)
  [hep-th/0309230].


\bibitem{Kastor:2006ti} 
  D.~Kastor and J.~Traschen,
  ``Stresses and Strains in the First Law for Kaluza-Klein Black Holes,''
  JHEP {\bf 0609}, 022 (2006)
  [hep-th/0607051].

\bibitem{Figueras:2011va} 
  P.~Figueras, J.~Lucietti and T.~Wiseman,
  ``Ricci solitons, Ricci flow, and strongly coupled CFT in the Schwarzschild Unruh or Boulware vacua,''
  Class.\ Quant.\ Grav.\  {\bf 28}, 215018 (2011)
  [arXiv:1104.4489 [hep-th]].

\bibitem{Haehl:2012tw} 
  F.~M.~Haehl,
  ``The Schwarzschild-Black String AdS Soliton: Instability and Holographic Heat Transport,''
  [arXiv:1210.5763 [hep-th]].

\bibitem{Fischetti:2012ps} 
  S.~Fischetti and D.~Marolf,
  ``Flowing Funnels: Heat sources for field theories and the $AdS_3$ dual of $CFT_2$ Hawking radiation,''
  Class.\ Quant.\ Grav.\  {\bf 29}, 105004 (2012)
  [arXiv:1202.5069 [hep-th]].


\end{thebibliography}
\end{document}